\documentclass[conference]{IEEEtran}
\IEEEoverridecommandlockouts

\usepackage{cite}
\usepackage{amsmath,amssymb,amsfonts}
\usepackage{mathtools}
\usepackage{graphicx}
\usepackage[table]{xcolor}
\usepackage{colortbl,hhline}
\usepackage{textcomp}
\usepackage{booktabs}
\usepackage{multirow}
\usepackage{tabularx}
\usepackage{microtype}

\usepackage{pgfplots}
\pgfplotsset{compat=1.18} 
\usepackage{tikz}
\usetikzlibrary{arrows.meta,positioning,matrix,backgrounds,shadows}

\usepackage{algorithm}
\usepackage[noend]{algorithmic}

\usepackage{bbm}
\usepackage{xspace}
\usepackage{listings}
\usepackage{tcolorbox}
\tcbuselibrary{listingsutf8}
\usepackage{varwidth}
\usepackage{float}
\usepackage{placeins}
\usepackage{multicol}
\usepackage{pdfpages}
\usepackage[left=1.62cm,right=1.62cm,top=1.9cm, columnsep=0.30in]{geometry}

\newcommand\sysname{\textsf{Sentinel}\xspace}

\begin{document}

\title{Enabling Trustworthy Federated Learning via Remote Attestation for Mitigating Byzantine Threats}

\author{\IEEEauthorblockN{
Chaoyu Zhang,
Heng Jin,
Shanghao Shi,
Hexuan Yu,
Sydney Johns,
Y. Thomas Hou, and 
Wenjing Lou}
\IEEEauthorblockA{Virginia Tech, VA, USA}
}

\maketitle

\begin{abstract}
Federated Learning (FL) has gained significant attention for its privacy-preserving capabilities, enabling distributed devices to collaboratively train a global model without sharing raw data. However, its distributed nature forces the central server to blindly trust the local training process and aggregate uncertain model updates, making it susceptible to Byzantine attacks from malicious participants, especially in mission-critical scenarios.
Detecting such attacks is challenging due to the diverse knowledge across clients, where variations in model updates may stem from benign factors, such as non-IID data, rather than adversarial behavior. Existing data-driven defenses struggle to distinguish malicious updates from natural variations, leading to high false positive rates and poor filtering performance.

To address this challenge, we propose \sysname, a remote attestation (RA)-based scheme for FL systems that regains client-side transparency and mitigates Byzantine attacks from a system security perspective. Our system employs code instrumentation to track control-flow and monitor critical variables in the local training process. Additionally, we utilize a trusted training recorder within a Trusted Execution Environment (TEE) to generate an attestation report, which is cryptographically signed and securely transmitted to the server. Upon verification, the server ensures that legitimate client training processes remain free from program behavior violation or data manipulation, allowing only trusted model updates to be aggregated into the global model. Experimental results on IoT devices demonstrate that \sysname ensures the trustworthiness of the local training integrity with low runtime and memory overhead.
\end{abstract}


\section{Introduction}\label{introduction}

Federated Learning is widely recognized for its privacy-preserving capabilities, enabling decentralized local clients to collaboratively train a global model without sharing raw data~\cite{konevcny2016federated1, mcmahan2017communication}. In mission-critical environments, FL enables collaborative intelligence by allowing heterogeneous nodes, such as satellites, UAVs, and maritime sensors, to jointly train models for disaster response and remote surveillance without requiring local data collection~\cite{zhao2024space, zhang2024state}. Similarly, underwater military operations benefit from FL through the efficient coordination of Internet of Underwater Things devices, which must operate under extreme communication and energy constraints~\cite{victor2023federated, zhang2023mindfl}. FL also plays a critical role in defending decentralized infrastructure, such as power and communication systems, by enabling distributed anomaly detection while preserving data locality and ensuring robustness against cyber threats~\cite{nist2023infra, wang2025feco, zhang2024hermes, zhang2025IDSSurvey}. Furthermore, in satellite-based remote sensing, FL supports fine-grained object recognition and target classification from geographically distributed training with proprietary datasets~\cite{chen2024free, zhang2025starcast}.

Despite its advantages, the distributed nature of FL introduces significant security vulnerabilities~\cite{chen2017distributed, so2020byzantine, guerraoui2018hidden, fang2020local, bhagoji2019analyzing, shi2025scale, Shi2025:CHASE:MedLeak}, such as Byzantine attacks, which are further amplified in mission-critical systems. Since the central server trusts remote clients to execute training correctly and use genuine data, compromised participants can exploit this trust. Byzantine nodes can degrade model performance or inject backdoors into the global model by directly manipulating their model updates (model poisoning attacks) or by using carefully crafted datasets (data poisoning attacks)~\cite{chen2017distributed, so2020byzantine, guerraoui2018hidden, fang2020local, bhagoji2019analyzing}.

\begin{figure}[t]
\centering
\includegraphics[width=\linewidth]{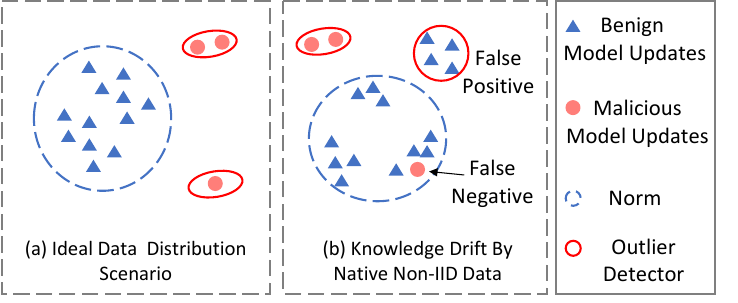}

\caption{Illustration of malicious model updates detection using a similarity-based mechanism. The detection is effective in an ideal IID scenario (case a) but is significantly impacted by inherent non-IID data (case b). In such cases, legitimate updates from clients with highly diverse data distributions or unique corner-case training samples may be misclassified.}

\label{noniid}
\vspace{-0.5cm}
\end{figure}

To address Byzantine attacks in FL, many studies adopt data-driven approaches to identify and filter potentially malicious model updates before aggregation, as illustrated in Fig.~\ref{noniid}. However, this remains difficult due to the inherent heterogeneity of local data, making it hard to distinguish benign drifts from adversarial manipulations. Traditional Byzantine-resilient methods rely on outlier detection, assuming malicious updates are separable in a common feature space~\cite{shen2016auror, fung2018mitigating, li2020learning}, as shown in Fig.~\ref{noniid}(a). Techniques include analyzing gradient angular distance~\cite{fung2018mitigating}, using low-dimensional embeddings~\cite{li2020learning}, and evaluating accuracy on synthetic data~\cite{zhao2019pdgan}. While effective under IID settings, these methods degrade under non-IID conditions~\cite{rieger2022deepsight, awan2021contra}, where legitimate updates from diverse clients may be misclassified as outliers, as shown in Fig.~\ref{noniid}(b). As a result, benign updates may be discarded or stealthy malicious ones accepted, reducing model performance or introducing backdoors~\cite{rieger2022deepsight, awan2021contra, briggs2020federated, wang2024flare, guerraoui2018hidden}.

To enhance trustworthiness in mission-critical FL, we propose \sysname, an RA-based framework that enhances client-side transparency from a system security perspective. RA allows a verifier to validate the integrity of software running on a remote device. In a typical RA protocol, the prover generates a cryptographic measurement, signed with a hardware-protected key, and sent to the verifier. This ensures the device runs the expected code and prevents report forgery. RA can be implemented using dedicated hardware (e.g., TEE) or CPU-based mechanisms. Our approach uses RA as a complementary solution to data-driven defenses, enabling verifiable client training in FL. The proposed framework comprises four key steps:

\textcircled{1} \textbf{Code Instrumentation:} To generate a trusted attestation report, our framework instruments the original training source code to track control-flow and monitor critical variables during local training. \textcircled{2} \textbf{Runtime Training Measurement:} To ensure measurement integrity, our design leverages a TEE as the root of trust. We introduce a trusted training recorder that dynamically records control-flow traces, monitors critical variable usage, and ensures the trustworthiness of execution reports sent to the task owner. \textcircled{3} \textbf{Attestation Report Generation:} To prevent forgery, the measurement engine signs execution data with its TEE signing key before transmission. The attestation report includes control-flow, critical variables usage, and a pre-issued challenge for the current FL iteration, with this signature. This attestation report is securely forwarded to the parameter server along with the model updates. \textcircled{4} \textbf{Training Verification:} The server verifies the local training process by analyzing execution paths and ensuring control-flow integrity (comparing recorded execution with expected logic) and data integrity (validating critical variable usage). If anomalies are detected, the corresponding model updates are flagged as potentially compromised and excluded from the current iteration's model aggregation.

By enforcing execution integrity on the client side, our approach avoids blindly trusting remote clients, reducing the attack surface and mitigating the potential Byzantine attacks that tamper with the training process. The key contributions of this work are summarized below:

\begin{itemize}  

\item \textbf{System-Oriented Byzantine Defense via RA:} We propose an RA-based framework for mission-critical or military FL that complements data-driven defenses by verifying local training integrity before model updates are accepted, mitigating both model and data poisoning attacks.

\item \textbf{Trusted Execution Measurement:} A trusted training recorder continuously monitors runtime execution, capturing control-flow traces and critical variable usage to detect adversarial manipulations and eliminate blind trust in clients.

\item \textbf{Verifiable Aggregation Without Heuristic Filtering:} Instead of relying on statistical filtering, our verification engine ensures only valid updates are aggregated by analyzing attestation reports for control and data integrity.

\item \textbf{Robustness With Low Overhead:} We prototype \sysname on Raspberry Pi V devices. Evaluation shows it effectively defends against runtime-level Byzantine attacks with minimal overhead in execution time and memory usage, significantly improving FL robustness.

\end{itemize}

\section{Related Work}\label{background}

As security and privacy of network communication systems evolve~\cite{du2023ucblocker, yuaaka, du2022mobile, li2023bijack, zhang2023mindfl, zhang2024hermes, zhang2025IDSSurvey, Yu2025closing, zhang2024state}, the RA is increasingly critical. RA has been applied to FL systems for various security purposes. DeepAttest~\cite{chen2019deepattest} embeds device-specific fingerprints in model weights, allowing TEEs to verify model legitimacy and prevent misuse. PPFL~\cite{mo2021ppfl} extends this by deploying TEEs on both clients and servers to protect updates from adversarial manipulation, mitigating passive attacks like data reconstruction and membership inference. SEAR~\cite{zhao2021sear} uses Intel SGX~\cite{intel_sgx} for secure aggregation, requiring clients to attest the server’s enclave before encrypting and submitting updates, preventing even honest-but-curious servers from recovering raw data. On the client side, an integrity-preserving FL scheme~\cite{zhang2020enabling} enforces protocol compliance by reproducing part of training inside a TEE and generating attestation reports, enabling the server to detect deviations like skipped updates or injected gradients. Additionally, several efforts focus on improving scheme efficiency~\cite{yu2021fpga, zhang2021high, fan2019gpu}.

Unlike prior work, which uses built-in TEE attestation services and tailors training to fit lightweight TEE constraints, we propose a customized RA scheme for FL that leverages the TEE as a root of trust for training process measurement. Our design follows the same TEE assumptions and usage as state-of-the-art RA frameworks~\cite{nunes2020apexRA1, abera2019diatRA3, sun2020oatRA4, wang2023ariRA7, Yu2025closing}.

\section{System Model and Threat Model}

\subsection{System Model}

We consider an FL system consisting of four main entities: the \textbf{task owner}, a set of \textbf{FL clients} $\{C_i\}|_{i\in[n]}$ (where $[n]\coloneqq\{1,2,...,n\}$), \textbf{trusted training recorders}, and \textbf{verification engines}. The task owner develops the FL system and implements the local training process. They define the model architecture, specify the local training protocol, and manage global model distribution and collection. The task owner also implements RA through code instrumentation in the local training process, enabling runtime measurement. Each FL client performs local training on its private dataset, collected during mission operation, while the training process is recorded to ensure integrity.  

In a typical FL training iteration $t$, the task owner distributes the global model weights $\theta^{(t)}$. Each client $C_i$ initializes local model weights $\mathbf{w}_i \in \mathcal{W} \subseteq \mathbb{R}^d$, where $\mathcal{W}$ is the parameter space and $d$ denotes the model dimensionality. The client then computes model updates $\delta_i^{(t)}$ in iteration $t$ in the form of weights or gradients for aggregation. Upon successful remote attestation of the local client, the task owner collects model updates $\{\delta_i^{(t)}\}_{i\in[v]}$ from selected clients and leverages the parameter server for model aggregation:
\[
\theta^{(t+1)} \leftarrow \frac{1}{|V|} \sum_{i\in \{C_i\}|_{i\in[V]}} \delta_i^{(t)}.
\]
The updated global model $\theta^{(t+1)}$ is then sent back to clients for the next FL iteration, and this process repeats until convergence. For measurement, each client is equipped with a trusted training recorder implemented within the TEE. For verification, the task owner collects attestation measurements from FL clients to validate the legitimacy of control-flow execution and the usage of critical variables.

\begin{figure*}[h]
\centering
\includegraphics[width=0.9\linewidth]{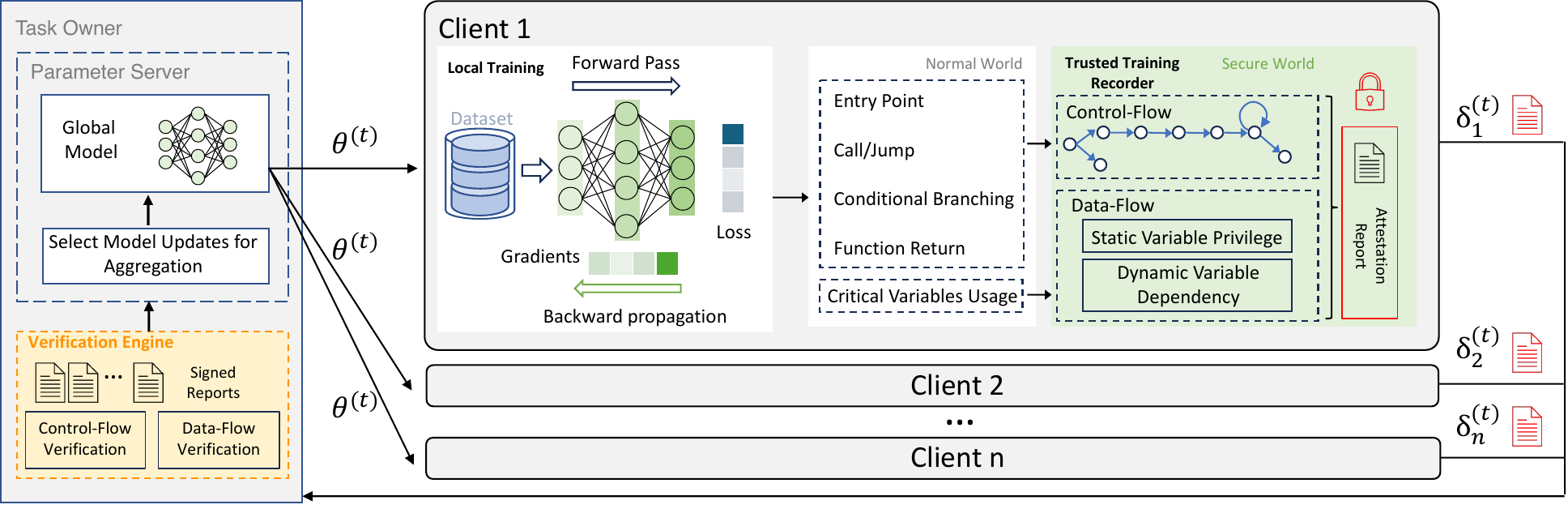}

\caption{\sysname enables trustworthy FL by recording the control-flow graph and critical variable usage through a trusted training recorder. The verification engine on the server side validates these reports and aggregates only verified updates, preventing blind trust in Byzantine attackers and ensuring transparency in the training process of remote devices.}
\label{deisgn}

\vspace{-0cm}

\end{figure*}

\begin{figure}[htbp!]
\centering
\includegraphics[width=\linewidth]{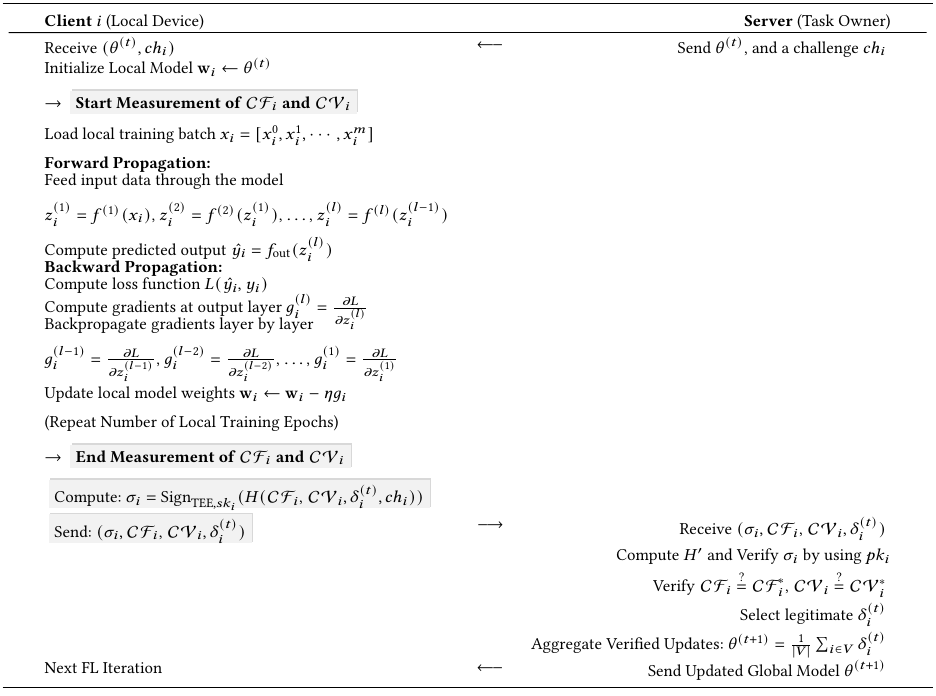}
\caption{\sysname Remote Attestation Protocol (\small Shadowed operations are executed within the TEE.)}
\label{Protocal}
\end{figure}

\subsection{Threat Model}

In \sysname, we consider the mission-critical FL scenario with the trustworthiness issue of local training as the key priority, where clients execute local training in a complex and untrusted environment, such as a battlefield.

The adversary in this work consists of Byzantine attackers who exploit software stack vulnerabilities to compromise the devices of FL clients deployed in military operations. These attackers violate the legitimate local training process to launch model or data poisoning attacks, aiming to degrade the global model's utility or inject backdoors.

We assume the task owner is a trusted entity that does not blindly trust remote training integrity and is responsible for verifying attestation reports and aggregating updates only from verified clients. The trusted computing base includes code instrumentation, the system enforcing it (e.g., privileged code managing the MMU or MPU~\cite{clements2018aces}), the TEE, and the underlying hardware. Specifically, the TEE, such as ARM TrustZone~\cite{arm_trustzone}, serves as the root of trust, ensuring secure collection, storage, and signing of attestation reports. The measurement engine is assumed to be tamper-resistant and correctly implemented. While our method does not directly defend against physical access attacks (e.g., sensor/input-based side channels or hardware fault injections), it may help identify such threats by detecting anomalies in execution patterns. A full defense against physical attacks requires further investigation.

\sysname is designed to detect remote program behavior violations, including control-flow hijacking and critical variable misuse during local training. It enhances transparency in the training process, eliminating blind trust in potentially compromised clients. \sysname complements existing data-driven approaches by adding a system-level layer of Byzantine resilience.

\section{Design of \sysname}\label{Design Details}

\subsection{Overview}

To enable trustworthy FL in mission-critical and military scenarios, we design an RA framework that enables verifiable local training, eliminating blind trust in remote clients. \sysname enforces control-flow integrity by recording forward-edge jumps/calls and return edges, ensuring clients follow the expected execution path. It also monitors critical variables, such as training data, model weights, gradients, and loss, to defend against data-only attacks.

\sysname workflow begins with the task owner, who implements the FL system, including server-side coordination and the on-device client application, while integrating the RA scheme into the workflow. To enable verifiable training on remote devices, the task owner \textit{instruments} the source code to monitor the control-flow graph $\mathcal{CF}_i$ and critical variable usage $\mathcal{CV}_i$, capturing both forward and backward edges to detect unauthorized re-directions. At the start of each training iteration $t$, client $C_i$ receives the global model weights $\theta^{(t)}$ and a unique attestation challenge $ch_i$, initializing local weights $\mathbf{w}_i \in \mathcal{W} \subseteq \mathbb{R}^d$. The client then performs local training and computes model updates $\delta_i^{(t)}$. A \textit{trusted training recorder}, implemented within a TEE, captures $\mathcal{CF}_i$ and $\mathcal{CV}_i$ during execution. At the end of the iteration, the client signs an \textit{attestation report}, bound to $ch_i$, using its TEE key $sk_i$, and sends the signed report along with $\delta_i^{(t)}$ to the task owner. The \textit{verification engine} authenticates the report, checks its integrity, and analyzes the control-flow for hijacking and variable usage for unauthorized modifications indicative of data-only attacks.

Only attestation reports that pass all verification checks are deemed valid, and their corresponding model updates are aggregated into the global model, computed as $\theta^{(t+1)}$. The updated global model is then redistributed to clients for the next training iteration. By integrating RA into the FL pipeline, our framework prevents compromised clients from contributing malicious updates. The use of a TEE ensures tamper-proof attestation, while private key signing of the report, including the unique challenge, prevents forgery. This guarantees that only verifiable and trustworthy contributions are included, significantly enhancing the transparency and robustness of the FL system.

\subsection{Remote Attestation Design}

We further illustrate how the RA scheme is integrated into the FL system in Fig.~\ref{deisgn}.

\textbf{Code Instrumentation:} To generate trusted attestation reports, our framework instruments the training application's source code to monitor control-flow execution and critical variable usage on remote devices.

We directly track the control-flow graph $\mathcal{CF}_i$ of each client $C_i$, including \textit{forward jumps/calls, return points, and conditional branches}, to ensure execution adheres to the expected flow and prevent unauthorized redirection. Unlike prior work, our design avoids using cryptographic hashes of return addresses or full control-flow traces for verification~\cite{abera2016cRA2}. This is motivated by the FL setting, where many clients perform short training sessions over a few epochs, and a central server must verify a high volume of reports.

In contrast to traditional cyber-physical systems, where frequent interrupts, callbacks, and dynamic control transfers make hash-based attestation beneficial~\cite{abera2016cRA2, sun2020oatRA4, wang2023ariRA7}, FL training is structured and iterative, with relatively static control-flow. Operations like forward/backward passes, gradient updates, and optimizer steps produce return instructions mainly at epoch boundaries or key computation points. Given this predictable execution, recording the full control-flow graph is both feasible and efficient, eliminating the overhead of hash chain reconstruction. This enables fast, deterministic verification and reliable detection of control-flow hijacking.

For critical variable usage $\mathcal{CV}_i$, we monitor two categories. The first includes static variables, such as training hyperparameters, e.g., learning rate, batch size, and the loaded dataset, which are configured as read-only before training begins. The second category includes dynamic variables, model weights, gradients, and loss, where we monitor dependencies and usage to detect tampering. All measurements are securely stored by a parallel trusted training recorder within the TEE, ensuring they cannot be manipulated or forged by Byzantine attackers.

\textbf{Runtime Training Measurement:} In each FL training iteration $t$, the task owner distributes global model weights $\theta^{(t)}$ to participating clients. Each client $C_i$ securely initializes its local model weights as $\mathbf{w}_i \leftarrow \theta^{(t)}$ in preparation for local training. This training phase consists of sequential control-flow and data-flow steps, where each step depends on the correctness of the preceding one.

Each client $C_i$ begins by loading a batch of local training data $x_i = [x_i^0, x_i^1, \cdots, x_i^m]$, where $m$ is the local batch size. The model processes the input across layers to compute a predicted output $\hat{y}_i$, completing the forward propagation. During this stage, the TEE records key training variables and operations, including the loading of input data and the layer-by-layer computation of intermediate representations $[z_i^{(1)}, z_i^{(2)}, \cdots, z_i^{(l)}]$, where $l$ denotes the number of layers.

Next, the client performs backward propagation to compute gradients $g_i$ and update parameters $\mathbf{w}_i$. The predicted output $\hat{y}_i$ is compared to ground-truth labels $y_i$ using a loss function $L(\hat{y}_i, y_i)$, which is back-propagated through the network via the chain rule. The TEE records gradient computations and relevant variables at each layer. Since each layer's gradients depend on the outputs of preceding layers, any tampering in backward propagation can be detected through inconsistency in later layers.

The final gradient $\nabla \mathbf{w}_i$ is used to update model weights as $\mathbf{w}_i \leftarrow \mathbf{w}_i - \eta \nabla \mathbf{w}_i$, where $\eta$ is the learning rate. This process repeats over multiple epochs on the client’s local dataset $\mathcal{D}_i$ until the model update $\delta_i^{(t)}$ is produced. Throughout the training, the trust engine securely stores control-flow measurements $\mathcal{CF}_i$ and critical variable usage $\mathcal{CV}_i$. In practice, these measurements are implemented using Secure Monitor Calls in ARM TrustZone~\cite{arm_trustzone}.

\textbf{Attestation Report Generation:} A malicious client may attempt to replay a previously valid model update from an earlier round to deceive the server. To prevent this, the attestation report must include a \textit{freshness guarantee}, ensuring that each attestation is uniquely bound to a specific training round. At the start of each round, the FL server generates a unique nonce (challenge $ch$) and distributes it to participating clients. Each client must incorporate its challenge $ch_i$ into the attestation report, which the server verifies before accepting the corresponding model update. This mechanism ensures that attestations are valid only for their intended round, effectively mitigating replay attacks. Additionally, timestamps can be embedded in the report using trusted hardware timers in ARM TrustZone, further strengthening temporal freshness guarantees. In our system, control-flow $\mathcal{CF}_i$ and critical variable usage $\mathcal{CV}_i$ are securely stored by the trusted training recorder. At the end of each FL iteration, clients sign the measurements and model updates, along with the challenge, using their TEE signing key: $\sigma_i = \text{Sign}_{\text{TEE}, sk_i}(H(\mathcal{CF}_i, \mathcal{CV}_i, \delta_i^{(t)}, ch_i)).$ The complete attestation report $(\sigma_i, \mathcal{CF}_i, \mathcal{CV}_i, \delta_i^{(t)})$ is then transmitted securely to the task owner via a standard secure communication channel for verification.

\textbf{Training Verification:} Upon receiving the attestation report $(\sigma_i, \mathcal{CF}_i, \mathcal{CV}_i, \delta_i^{(t)})$, the server first verifies its authenticity and integrity. It then validates both the control-flow integrity $\mathcal{CF}_i$ and critical variable usage $\mathcal{CV}_i$.

For control-flow verification, the server analyzes the execution trace and checks its consistency with the expected control-flow graph $\mathcal{CF}^*_i$ of the training process. The verifier iterates through the trace, validating branch conditions, jump/call targets, and return points. Instead of relying on hash-based methods, our design records explicit forward and return points, allowing the server to detect inconsistencies in call/return sequences indicative of control-flow hijacking. Verification halts immediately upon detecting a mismatch, minimizing computational overhead.

To verify critical variable usage $\mathcal{CV}_i$, the server ensures that static variables, such as hyperparameters and dataset configurations, remain read-only throughout training. Reported values are compared against expected usage $\mathcal{CV}^*_i$, and any unauthorized modification leads to rejection. For dynamic variables, including model weights, gradients, and loss values, verification checks whether updates follow valid data dependencies. For instance, model weights must be updated via optimizer steps, gradients must result from backpropagation, and loss values must derive from valid forward-pass computations.

If any anomaly is detected, such as inconsistent propagation sequences or unauthorized variable usage, the corresponding model updates are flagged as malicious and excluded from aggregation. Only updates that conform to both control-flow and data-flow constraints are accepted, preventing Byzantine clients from injecting manipulated results into the global model. The detailed protocol of \sysname is illustrated in Fig.~\ref{Protocal}.

\section{Evaluation}

\begin{figure}[h]
\centering
\includegraphics[width=\linewidth]{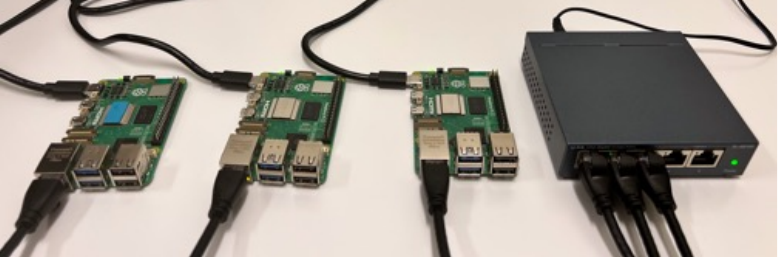}

\caption{Testbed: Raspberry Pi V as the Training Devices for Local Clients in RA Implementation.}

\vspace{-0.1cm}

\label{testbed}
\end{figure}

\begin{table*}[t]
    \centering
    \caption{Targeted Byzantine Resilience Performance in FL: Evaluated by Global Model Test Accuracy and Attack Success Rate Triggered by Crafted Inputs under IID and Non-IID Settings.}

    \label{tab:target}
    \begin{tabular}{lcccccccc}
        \toprule
        \multirow{2}{*}{\textbf{Attack}} & \multirow{2}{*}{\textbf{Dataset}} & \textbf{FedAvg} & \textbf{Krum} & \textbf{Coomed} & \textbf{TrimmedMean} & \textbf{Bulyan} & \textbf{FLTrust} & \textbf{\sysname} \\ 
        
        \cmidrule(lr){3-9}
        & & Acc/ASR & Acc/ASR & Acc/ASR & Acc/ASR & Acc/ASR & Acc/ASR & Acc/ASR \\ \midrule
       
        \multirow{4}{*}{T-CF-krum} 
        & FMNIST (IID) & 91.7/8.4 & 88.0/97.8 & 91.8/44.1 & 91.6/43.9 & 91.4/71.2 & 91.6/39.6 & 91.3/0.0 \\
        & FMNIST (non-IID) & 87.2/18.4 & 83.5/99.5 & 87.0/59.0 & 87.3/55.0 & 87.2/81.0 & 88.5/49.5 & 89.5/0.0 \\
        & CIFAR-10 (IID) & 68.5/25.9 & 48.9/99.5 & 67.9/44.3 & 68.1/35.7 & 67.3/66.0 & 66.9/1.7 & 66.6/0.0 \\
        & CIFAR-10 (non-IID) & 63.5/38.5 & 45.0/99.9 & 63.0/59.5 & 63.0/44.0 & 62.5/76.0 & 64.0/28.3 & 64.5/0.0 \\ \midrule
        
        \multirow{4}{*}{T-DO-krum} 
        & FMNIST (IID) & 91.7/57.8 & 87.8/98.7 & 91.9/77.9 & 91.5/67.5 & 91.4/85.2 & 91.6/0.1 & 91.4/0.0 \\
        & FMNIST (non-IID) & 87.2/70.5 & 83.0/99.8 & 87.1/89.5 & 87.0/80.0 & 86.8/93.5 & 88.5/5.0 & 89.2/0.0 \\
        & CIFAR-10 (IID) & 68.2/55.3 & 50.5/97.9 & 67.7/81.2 & 67.5/54.8 & 66.3/89.1 & 66.7/5.3 & 65.7/0.0 \\
        & CIFAR-10 (non-IID) & 63.2/67.8 & 45.5/99.9 & 63.0/93.0 & 62.8/69.0 & 62.0/95.0 & 63.2/26.5 & 63.0/0.0 \\ 
        \bottomrule
    \end{tabular}
\end{table*}

\begin{table*}[t]
    \centering
    \caption{Untargeted Byzantine Resilience Performance in FL: Evaluated Based on Global Model Test Accuracy under IID and Non-IID Settings.}

    \label{tab:untarget}
    \begin{tabular}{lcccccccc}
        \toprule
        \textbf{Attack} & \textbf{Dataset} & \textbf{FedAvg} & \textbf{Krum} & \textbf{Coomed} & \textbf{TrimmedMean} & \textbf{Bulyan} & \textbf{FLTrust} & \textbf{\sysname} \\ \midrule

    \multirow{4}{*}{UNT-CF-krum} 
    & FMNIST (IID) & 59.4 & 7.5 & 79.8 & 89.5 & 86.5 & 90.8 & 91.0 \\
    & FMNIST (non-IID) & 56.0 & 5.0 & 75.5 & 86.0 & 83.5 & 88.5 & 89.2 \\
    & CIFAR-10 (IID) & 59.1 & 11.0 & 68.5 & 68.4 & 68.3 & 66.4 & 67.2 \\ 
    & CIFAR-10 (non-IID) & 57.5 & 6.5 & 58.0 & 58.0 & 60.0 & 63.5 & 64.8 \\ 
    \midrule
    
    \multirow{4}{*}{UNT-DO-krum} 
    & FMNIST (IID) & 87.3 & 88.3 & 80.5 & 60.7 & 89.6 & 90.9 & 91.1 \\
    & FMNIST (non-IID) & 84.0 & 85.5 & 77.0 & 58.0 & 86.0 & 89.0 & 90.5 \\
    & CIFAR-10 (IID) & 58.8 & 88.5 & 69.0 & 61.5 & 68.6 & 66.8 & 67.4 \\ 
    & CIFAR-10 (non-IID) & 56.0 & 84.0 & 65.5 & 49.5 & 60.0 & 61.5 & 65.0 \\
    \bottomrule

\end{tabular}
\end{table*}

\begin{table*}[t]
    \centering
    \caption{System overhead for different ML tasks with RA, evaluated on a Raspberry Pi V. Each configuration was trained on 64, 128, and 256 samples for one epoch. Results are averaged over 64 runs.}

    \label{tab:attestation_overhead}
    \includegraphics[width=0.9\linewidth]{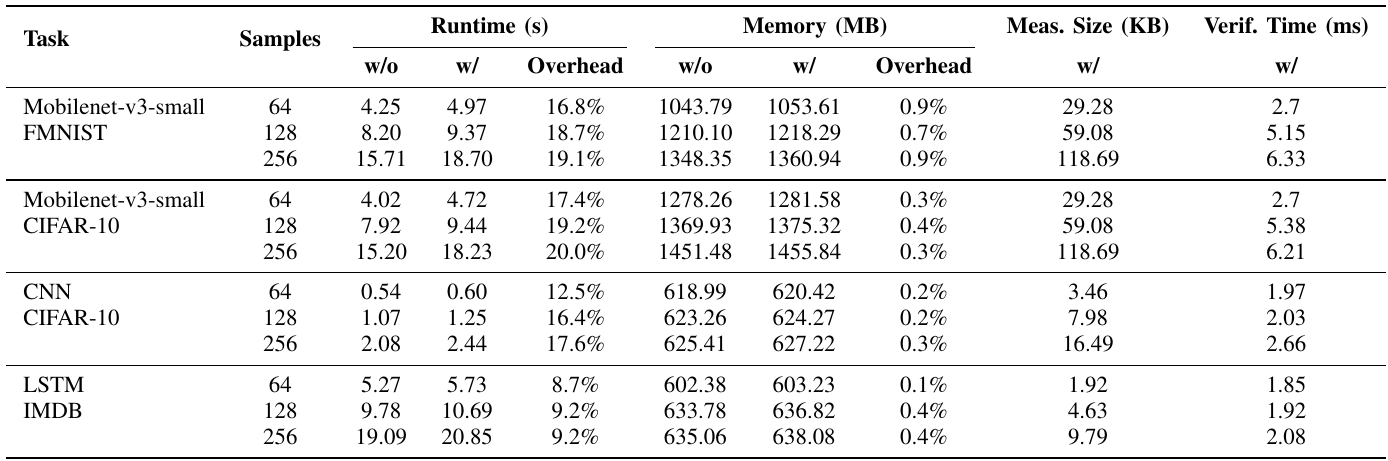}
    
    \vspace{-0cm}

\end{table*}

\subsection{Experimental Setup}
In our FL system, a parameter server coordinates 100 clients, each training a local model on private datasets initialized with global model weights. Datasets are partitioned as follows: (1) IID, where each client receives equal data covering all classes, and (2) non-IID, where each client receives equal data but only half of the classes randomly. Clients train locally for five epochs using Adam (learning rate 0.001) before submitting updates, with a total of 30 FL iterations. We evaluate on FMNIST~\cite{xiao2017fashion} and CIFAR-10~\cite{krizhevsky2009learning}. Experiments are conducted on a server with an Intel Core i7 CPU, an NVIDIA GeForce RTX 3080 GPU, Ubuntu 22.04, and three Raspberry Pi V devices as simulated clients.

We compare our approach with \textbf{FedAvg}~\cite{mcmahan2017communication} and several defense baselines, including \textbf{Krum}~\cite{blanchard2017machine}, \textbf{Coomed}~\cite{yin2018byzantine}, \textbf{TrimmedMean}~\cite{yin2018byzantine}, \textbf{Bulyan}~\cite{guerraoui2018hidden}, and \textbf{FLTrust}~\cite{cao2021fltrust}. To evaluate, we use \textbf{control-flow hijacking} and \textbf{data-only attacks} that compromise local training and alter model updates. Specifically, we implement Krum-Backdoor (targeted)~\cite{bhagoji2019analyzing} and Krum-Untargeted~\cite{fang2020local} attacks.

\begin{figure}[t]
\centering
\includegraphics[width=\linewidth]{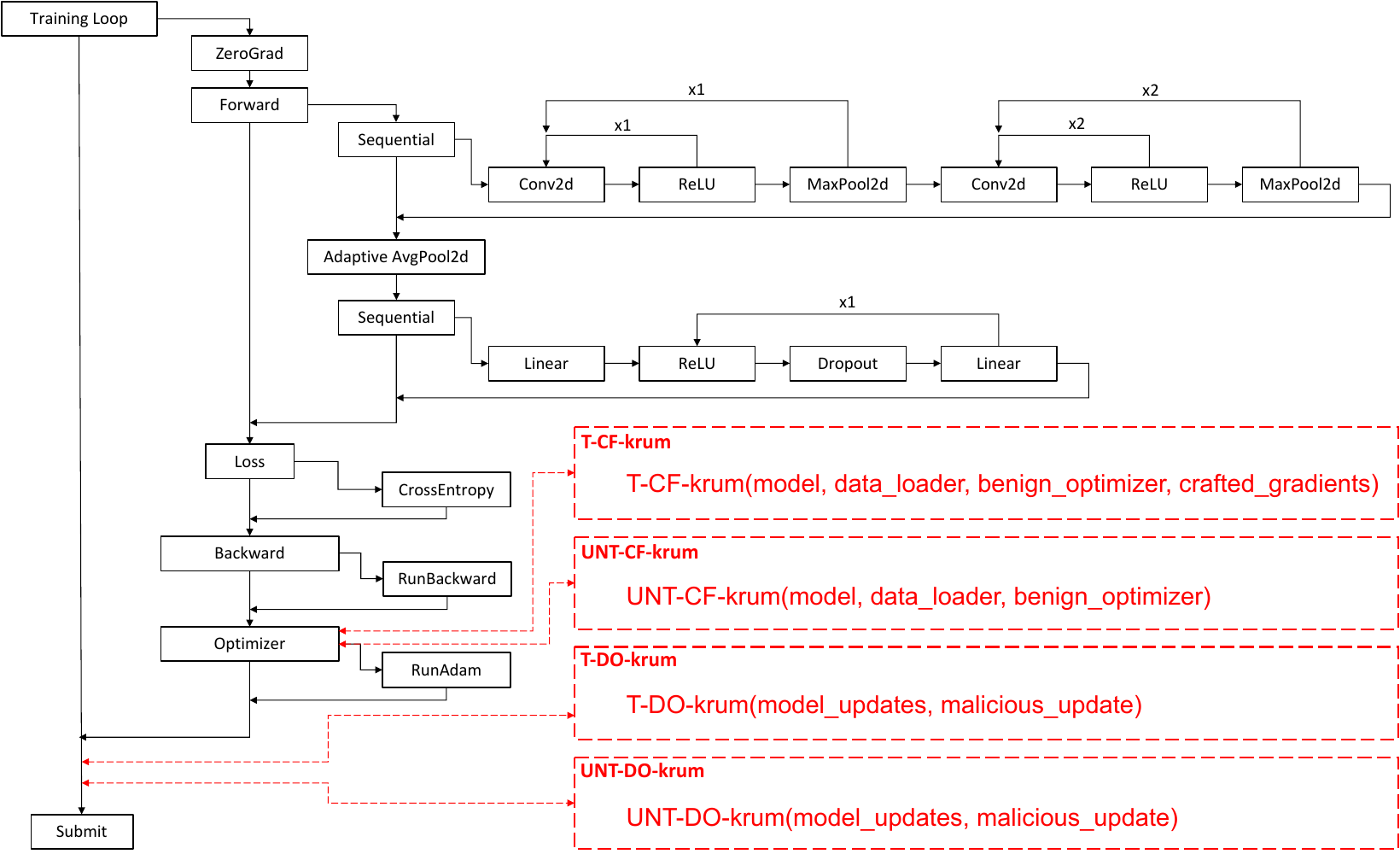}

\caption{We consider four attack strategies: \textbf{T-CF-Krum} and \textbf{UNT-CF-Krum}, which use control-flow hijacking to redirect execution to a malicious optimizer, and \textbf{T-DO-Krum} and \textbf{UNT-DO-Krum}, which perform data-only attacks to replace benign ones.
}

\label{CFA}
\end{figure}

To test \sysname, we design targeted and untargeted control-flow attacks, \textbf{T-CF-Krum} and \textbf{UNT-CF-Krum}, which redirect execution to a malicious optimizer, and data-only attacks, \textbf{T-DO-Krum} and \textbf{UNT-DO-Krum}, which replace benign updates with adversarial ones (Fig.~\ref{CFA}).

We evaluate defenses using model accuracy (Acc) and attack success rate (ASR). For targeted attacks, ASR is the proportion of test inputs with a trigger misclassified as the target label. For untargeted attacks, where the goal is model degradation, Acc is used as the sole metric.

\subsection{Defense Effectiveness Results and Analysis}

While some data-driven defenses can detect adversarial updates under IID conditions, they struggle in non-IID settings due to diverse client knowledge. Variations in model updates may result from legitimate drift rather than adversarial manipulation, causing poor detection and weak mitigation.

Results of targeted attacks (Tab.~\ref{tab:target}) confirm this limitation. In IID settings, defenses such as Krum and Bulyan fail to reduce ASR, and their performance degrades further in non-IID cases, where ASR remains high. Since these aggregation rules rely on anomaly detection, they cannot distinguish adversarial updates from natural variations in heterogeneous clients. Similarly, under untargeted attacks (Tab.~\ref{tab:untarget}), data-driven methods misclassify legitimate updates as malicious, reducing accuracy in non-IID scenarios and leaving models vulnerable. This highlights their reliance on fragile statistical heuristics.

In contrast, \sysname\ integrates robust aggregation with a principled defense that separates adversarial behavior from natural heterogeneity. Our results show \sysname\ nullifies targeted attacks (ASR = 0\% in all cases) and achieves the highest accuracy under untargeted attacks, demonstrating clear advantages over conventional defenses.

\subsection{System Overhead Analysis}

To evaluate the feasibility and overhead of integrating remote attestation into FL, we deploy Raspberry Pi V devices with Arm Cortex-A76 processors as FL clients to build a local training testbed (see Fig.~\ref{testbed}). Each device simulates local training while running a parallel trusted training recorder. Code instrumentation in the source interpreter traces critical events, monitors stack frames, and inspects instructions during key workflows. Critical variables are verified before important steps and continuously monitored, with modifying instructions recorded. Each FL client trains on a small private dataset, reflecting real-world, privacy-sensitive on-device scenarios.

For consistency, models are trained with fixed batch sizes of 64, 128, and 256 for a single epoch, ensuring a controlled environment to measure attestation impact. All experiments are repeated 64 times, and average results are reported. We evaluate multiple ML tasks: CNNs and MobileNetV3 on FMNIST and CIFAR-10 for image classification, and an LSTM on IMDB for text classification. These cover workloads from lightweight CNNs to sequential models, providing a comprehensive view of attestation overhead across architectures. Metrics include runtime, memory usage, attestation report size, and verification latency (Tab.~\ref{tab:attestation_overhead}).

\textit{Runtime Overhead:} Remote attestation introduces moderate runtime increases, mainly from real-time measurement and trusted recording. More complex models with larger parameter sizes incur slightly higher overhead. \textit{Memory Overhead:} Memory usage shows minimal growth, as the attestation mechanism relies on lightweight monitoring. \textit{Measurement Size and Verification Latency:} Measurement size depends on model control-flow complexity and tracked variables, with deeper models producing larger records. Verification time, however, remains low across all tasks, avoiding bottlenecks in FL deployments.

\section{Conclusion}

In this work, we propose \sysname, a remote attestation-based framework to enhance the security and transparency of FL training on remote devices. Unlike data-driven Byzantine-resilient approaches, \sysname enforces system-level integrity checks to mitigate model poisoning and data poisoning attacks. By leveraging a TEE-based trusted training recorder, our design ensures that FL clients follow the expected training execution flow and do not manipulate critical variables. Experimental results demonstrate that \sysname effectively detects adversarial manipulations while maintaining a low overhead in execution time and memory usage, making it practical for mission-critical devices for FL deployments.

\section*{Acknowledgment}
This work was supported in part by the Office of Naval Research under grants N00014-24-1-2730, and the National Science Foundation under grants 2312447, 2247560, 2154929, and 2235232.

\bibliographystyle{ieeetr}
\bibliography{IEEEabrv,ref}


\end{document}